\def\psiket{|\psi\rangle}
\def\up{|\!\!\uparrow\rangle}
\def\down{|\!\!\downarrow\rangle}
\def\smileface{\raisebox{-2pt}{$\ddot\smile$}}
\def\frownface{\raisebox{-2pt}{$\ddot\frown$}}

\def\noobs{|\,\ddot{\_}\,\rangle}
\def\noobs{|\,\ddot{\raisebox{-2pt}{-}}\,\rangle}
\def\upobs{|\smileface\rangle}

\def\downobs{|\frownface\rangle}
\def\tensormult{\otimes} 

\def\kg{{\rm kg}}

\def\rn{}
\def\nn#1 #2{#2. #1}				
\def\nnn#1 #2 #3{#2. #3. #1}			
\def\nnnn#1 #2 #3 #4{#2. #3. #4 #1}		
\def\nnnnn#1 #2 #3 #4 #5{#2. #3. #4 #5. #1}	
\def\dualand{ and\hbox{ }}				
\def\multiand{, and\hbox{ }}				
\def\rf#1;#2;#3;#4;#5 {{\frenchspacing\par\rn#1, #3 {\bf #4}, #5 (#2). \par}}
\def\rg#1;#2;#3;#4;#5;#6 {{\frenchspacing\par\rn#1, #3 {\bf #4}, #5 (#2). \par}}
\def\rfproc#1;#2;#3;#4;#5;#6 {{\frenchspacing\par\rn#1 #2, in {\it #3}, ed. #4 (#5: #6)\par}}
\def\rfprocp#1;#2;#3;#4;#5;#6;#7 {{\frenchspacing\par\rn#1 #2, in {\it #3}, ed. #4 (#5: #6), p#7\par}}

\def\rg#1;#2;#3;#4;#5;#6 {\par\rn#1 #2, {\it #3}, {\bf #4}, #5 (``#6'') \par}
\def\rf#1;#2;#3;#4;#5 {\par\rn#1, {\it #3}, {\bf #4}, #5 (#2)\par}
\def\rfbook#1;#2;#3;#4;#5 {{\frenchspacing\par\rn#1, {\it #3} (#4: #5, #2)\par}}
\def\rfproc#1;#2;#3;#4;#5;#6 {{\frenchspacing\par\rn#1 #2, in {\it #3}, ed. #4 (#5: #6)\par}}
\def\rfprocp#1;#2;#3;#4;#5;#6;#7 {{\frenchspacing\par\rn#1 #2, in {\it #3}, ed. #4 (#5: #6), p#7\par}}
\def\rfprep#1;#2;#3 {{\par\frenchspacing\rn#1, #3 (#2)\par}}
\def\rfprepp#1;#2;#3 {{\par\rn#1 #2, #3\par}}

\def\ie{{\frenchspacing\it i.e.}}
\def\eg{{\frenchspacing\it e.g.}}
\def\etc{{\frenchspacing\it etc.}}

\def\beq#1{\begin{equation}\label{#1}}
\def\eeq{\end{equation}}
\def\beqa#1{\begin{eqnarray}\label{#1}}
\def\eeqa{\end{eqnarray}}
\def\eq#1{equation~(\ref{#1})}

\def\fig#1{Figure~\ref{#1}}

\def\Sec#1{Section~\ref{#1}}
\def\Sec#1{Section~\ref{#1}}

\def\spose#1{\hbox to 0pt{#1\hss}}
\def\simlt{\mathrel{\spose{\lower 3pt\hbox{$\mathchar"218$}}
     \raise 2.0pt\hbox{$\mathchar"13C$}}}
\def\simgt{\mathrel{\spose{\lower 3pt\hbox{$\mathchar"218$}}
     \raise 2.0pt\hbox{$\mathchar"13E$}}}
\def\simpropto{\mathrel{\spose{\lower 3pt\hbox{$\mathchar"218$}}
     \raise 2.0pt\hbox{$\propto$}}}

\def\ed{\end{document}}

\documentclass[twocolumn,amsmath,nofootinbib]{revtex4} 
\usepackage{amsfonts,amsbsy,epsf} 
\begin{document}







\title{Many Worlds in Context\footnote{To appear in {\it ``Many Worlds? Everett, Quantum Theory and Reality"}, S.~Saunders, J.~Barrett, A.~Kent \& D.~Wallace (eds), Oxford Univ.~Press}}

\author{Max Tegmark}

\address{Dept.~of Physics, Massachusetts Institute of Technology, Cambridge, MA 02139; tegmark@mit.edu}

\begin{abstract}
Everett's Many-Worlds Interpretation of quantum mechanics is discussed in the context of other 
physics disputes and other proposed kinds of parallel universes.
We find that only a small fraction of the usual objections to Everett's theory are specific to quantum mechanics,
and that all of the most controversial issues crop up also in settings that have nothing to do with quantum mechanics.
\end{abstract}

\date{Submitted August 3 2008, revised February 14 2010}


  
\maketitle



\setcounter{footnote}{0}

\def\thetamin{\theta_{\rm min}}
\def\thetamax{\theta_{\rm max}}

\section{Introduction}

There is now great interest in Everett's Many-Worlds Interpretation of quantum mechanics and the controversy 
surrounding it.\footnote{The controversy shows no sign of abating, as evidenced by the results of the following highly unscientific poll
carried out by the author at the Perimeter Institute ``Everett@50'' Conference 9/22-07:
\itemsep0mm
\begin{enumerate} 
\item {\it Do you believe that new physics violating the Schr{\"o}dinger equation will make large quantum computers impossible?} (4 Yes/ 29 No/11 Undecided)
\item {\it Do you believe that all isolated systems obey the Schr{\"o}dinger equation (evolve unitarily)?} (17 Yes/10 No/20 Undecided)
\item {\it Which interpretation of quantum mechanics is closest to your own?}
\begin{itemize}
  \itemsep0mm
  \item 2 Copenhagen or consistent histories (including postulate of explicit collapse)
  \item 5 Modified dynamics (Schršdinger equation modified to give explicit collapse) 
  \item 19 Many worlds/consistent histories (no collapse) 
  \item  2 Bohm 
  \item  1.5 Modal 
  \item  22.5 None of the above/undecided
\end{itemize}
\item {\it Do you feel comfortable saying that Everettian parallel universes are as real as our universe?} (14 Yes/26 No/8 Undecided)
\end{enumerate}
}
A key reason for this is undoubtedly that it connects with some of our deepest questions about the nature of reality. How large is physical reality? Are there parallel universes? Is there fundamental randomness in nature? 

The goal of this article is to place both Everett's theory and the standard objections to it in context. We will review how Everett's Many Worlds may constitute merely one out of four different levels of parallel universes, the rest of which have little to do with quantum mechanics. 
We will also analyze the many objections to Everett's theory listed in Table 1, concluding that most of them are not specific to quantum mechanics. 
By better understanding this context, quantum physicists can hopefully avoid
reinventing many wheels that have been analyzed in detail in other areas of physics or philosophy, and focus their efforts on those remaining aspects of Everett's theory that are 
uniquely quantum-mechanical. This is not to say that the issues in Table 1 with a ``No'' in the {\it QM-specific} column are necessarily unimportant --- merely that it is unfair to blame Hugh Everett for them or to use them as evidence against his theory alone.

Rather than discuss these objections one by one in the order they appear in Table 1, this article is structured as a survey of multiverse theories, addressing the objections in their natural context. We then return to Table 1 and summarize our conclusions in \Sec{DiscSec}.


\begin{table*}[t!!]
{\it {\bf Table~1.} Most common worries about Everett's many-worlds interpretation are not
specific to quantum mechanics.}
{ \footnotesize
\begin{tabular}{|rlcl|}
\hline
  &Worry								&QM-specific?	&Counterexamples/resolution\\
\hline
 1&Popper worry: falsifiable?						&No	&General relativity, inflation\\
 2&Occam worry: parallel universes feel wasteful			&No	&Level I \& 2 multiverses\\
 3&Aristotle worry: math mere approximation				&No	&Linked to expernal reality hypothesis\\
 4&Uncertainty worry: How can omniscience allow uncertainty?		&No	&Occurs whenever observer ensemble\\
 5&How derive probabilities from deterministic theory?			&No	&Occurs whenever observer cloning\\
 6&Unequal probability worry: Why square the amplitudes?		&Yes	&Comes from Hilbert space structure\\
 7&$\rho$ worry: describes world or my knowledge of it?			&Partly	&Can describe both\\
 8&How judge evidence for/against such a theory?			&No	&Classical statistical mechanics\\
 9&Word worry: What do we mean by ``exist", ``real", ``is", etc?	&No	&Level I \& II multiverses\\
10&Invisibility worry: Why can't we detect the parallel worlds?		&Yes	&Solved by decoherence\\
11&Basis worry: What selects preferred basis?				&Yes	&Solved by decoherence\\
12&Weirdness worry							&No	&Electric fields, black holes, Levels I \& II\\
\hline
\end{tabular}
}
\end{table*}

\subsection{The MWI: what it is and what it isn't}
\label{MWIdefSec}

Let us first spell out what we mean by the Many Worlds Interpretation (MWI) 
Much of the early criticism of the MWI was based on 
confusion as to what it meant. Here we grant Everett 
the final say in how the MWI is defined,  since he did 
after all invent it \cite{Everett57,EverettBook}, and take it to consist of 
the following postulate alone:

\begin{itemize}
\item {\bf EVERETT POSTULATE:}\\
{\it All isolated systems evolve according to the Schr\"odinger equation
${d\over dt}\psiket=-{i\over\hbar}H\psiket.$}
\end{itemize}
More succinctly, ``physics is unitary".
 Although this postulate sounds rather innocent, it has far-reaching 
implications:
\begin{enumerate} 
\item {\bf Corollary 1:} the entire Universe evolves 
according to the Schr\"odinger equation, since it is 
by definition an isolated system. 
\item {\bf Corollary 2:} when a superposition state is observed, there can be no definite outcome 
(wavefunction collapse), since this would violate the Everett postulate.
\end{enumerate}
Because of corollary 1, ``universally valid quantum mechanics'' 
is often used as a synonym for the MWI. What is to be 
considered ``classical'' is therefore {\it not} specified axiomatically 
(put in by hand) in the MWI --- rather, it can be derived 
from the Hamiltonian dynamics, by computing decoherence rates.

How does corollary 2 follow? 
Consider a measurement of a spin 1/2 system (a silver atom, say)
where the states 
``up'' and ``down''
along the $z$ axis are denoted $\up$ and $\down$. 
Assuming that the observer will get happy if she measures spin up, 
we let $\noobs$, $\upobs$ and $\downobs$ denote the states of 
the observer before the measurement, after perceiving 
spin up and after perceiving spin down, respectively.
If the measurement is to be described by a unitary 
Schr\"odinger time evolution operator $U=e^{-iH\tau/\hbar}$ applied to the
total system, then $U$ must clearly satisfy  
\beq{Ueq}
U\up\tensormult\noobs = \up\tensormult\upobs
\quad\hbox{and}\quad
U\down\tensormult\noobs = \down\tensormult\downobs.
\eeq
Therefore if the atom is originally in a superposition
\hbox{$\alpha\up+\beta\down$}, then the Everett postulate implies that
the state resulting after the observer has interacted
with the atom is
\beq{SplitEq}
U(\alpha\up+\beta\down)\tensormult\noobs = 
\alpha\up\tensormult\upobs + \beta\down\tensormult\downobs.
\eeq
In other words, the outcome is not $\up\tensormult\upobs$
or $\down\tensormult\downobs$ with some probabilities, merely 
these two states in superposition.
Very few physicists have actually read Everett's original 137-page Ph.D. thesis 
(reprinted in \cite{EverettBook}), 
which has lead to a common misconception that it 
contains a second postulate along the following lines:
\begin{itemize}
\item What Everett does {\bf NOT} postulate:\\
{\it At certain magic instances, the world undergoes
some sort of metaphysical ``split'' into 
two branches that subsequently never interact.}
\end{itemize}
This is not only a misrepresentation of the MWI, but also inconsistent 
with the Everett postulate, since the subsequent time evolution could in 
principle make the two terms in 
\eq{SplitEq} interfere. 
According to the MWI, there is, was and always will be
only one wavefunction, and 
only decoherence calculations, not postulates, 
can tell us when it is a good approximation to treat two terms 
as non-interacting.

\subsection{Many worlds galore}

Parallel universes are now all the rage, cropping up in books, movies and 
even jokes: ``You passed your exam in many parallel universes --- but not this one."
However, they are as controversial as they are popular, and it is important to ask whether they are 
within the purview of science, or merely silly speculation. 
They are also a source of confusion, since many forget to distinguish between 
different types of parallel universes that have been proposed\cite{multiverse,multiverse4wheeler}.

The farthest you can observe is the 
distance that light has been able to travel during the 14 billion years since the big-bang 
expansion began. The most distant visible objects are now about $4\times 10^{26}$ meters 
away\footnote{After emitting 
the light that is now reaching us, 
the most distant things we can see have receded because of the cosmic expansion,
and are now about about 40 billion light years away.
},
and a sphere of this radius defines our observable universe, also called our {\it Hubble volume}, our 
{\it horizon volume} or simply our universe. 
Below I survey physics theories involving parallel universes, which 
form a natural four-level hierarchy of multiverses (Figure 1) allowing progressively greater diversity.
\begin{itemize}
\item {\bf Level I:} A generic prediction of cosmological inflation is an infinite ``ergodic'' space,
which contains Hubble volumes realizing all initial conditions --- including
an identical copy of you about $10^{10^{29}}$m away.\\
\item {\bf Level II:} Given the {\it fundamental} laws of physics that physicists one day hope to capture with equations on a T-shirt, 
different regions of space can exhibit different
{\it effective} laws of physics (physical constants, dimensionality, particle content, \etc) corresponding
to different local minima in a landscape of possibilities. 
\item {\bf Level III:} In Everett's unitary quantum mechanics, other branches of the wavefunction 
add nothing qualitatively new, which is ironic given
that this level has historically been the most controversial.
\item {\bf Level IV:} Other mathematical structures give different fundamental equations of physics for that T-shirt.
\end{itemize}
The key question is therefore not whether there is a multiverse (since Level I is the rather uncontroversial 
cosmological concordance model), but rather how many levels it has.

Below we will discuss at length the issue of evidence and whether this is science or philosophy.
For now, the key point to remember is that
{\it parallel universes are not a theory, but a prediction of certain theories}.
The Popper worry listed in Table 1 is the question of whether Everett's theory is falsifiable.
For a theory to be falsifiable, we need not be able to observe and test all its predictions, merely at least one of them.
Consider the following analogy:
\begin{center}
\begin{tabular}{|l|l|}
\hline
General Relativity      &Black hole interiors\\
\hline
Inflation               &Level I parallel universes\\
\hline
Unitary quantum mechanics&Level III parallel universes\\
\hline
\end{tabular}
\end{center}
Because Einstein's theory of General Relativity has successfully predicted many things that we {\it can} observe, we also take seriously its predictions for things we
cannot observe, \eg, that space continues inside black hole event horizons and that (contrary to early misconceptions) nothing funny happens
right at the horizon.
Likewise, successful predictions of the theories of cosmological inflation and 
unitary\footnote{As described below, 
the mathematically simplest version of quantum mechanics is ``unitary'', lacking the 
controversial process known as wavefunction collapse.}
quantum mechanics have made some scientists take more seriously their other predictions, 
including various types of parallel universes.

Let us conclude with two cautionary remarks before delving into the details.
H{\"u}bris and lack of imagination have repeatedly caused us humans to underestimate the vastness of the physical world,
and dismissing things merely because we cannot observe them from our vantage point is reminiscent of the ostrich with its head in the sand.
Moreover, recent theoretical insights have indicated that Nature may be tricking us.
Einstein taught us that space is not merely a boring static void, but a dynamic entity that can stretch (the expanding universe), 
vibrate (gravitational waves) and curve (gravity). Searches for a unified theory also suggest that space can ``freeze'', 
transitioning between different phases in a landscape of possibilities just like water can be
solid, liquid or gas. In different phases, effective laws of physics (particles, symmetries, \etc.) could differ.
A fish never leaving the ocean might mistakenly conclude that the properties of water are universal, not realizing that there is also ice and steam.
We may be smarter than fish, but could be similarly fooled: cosmological inflation has the deceptive property of stretching a small patch of space in a particular phase 
so that it fills our entire observable universe, potentially tricking us into misinterpreting our local conditions for the universal laws that should go on that T-shirt.

\clearpage
\begin{figure}[tbp]
\vskip0cm
\label{ZoomFig}
\centerline{{\vbox{\hglue-0.5cm\epsfxsize=18.4cm\epsfbox{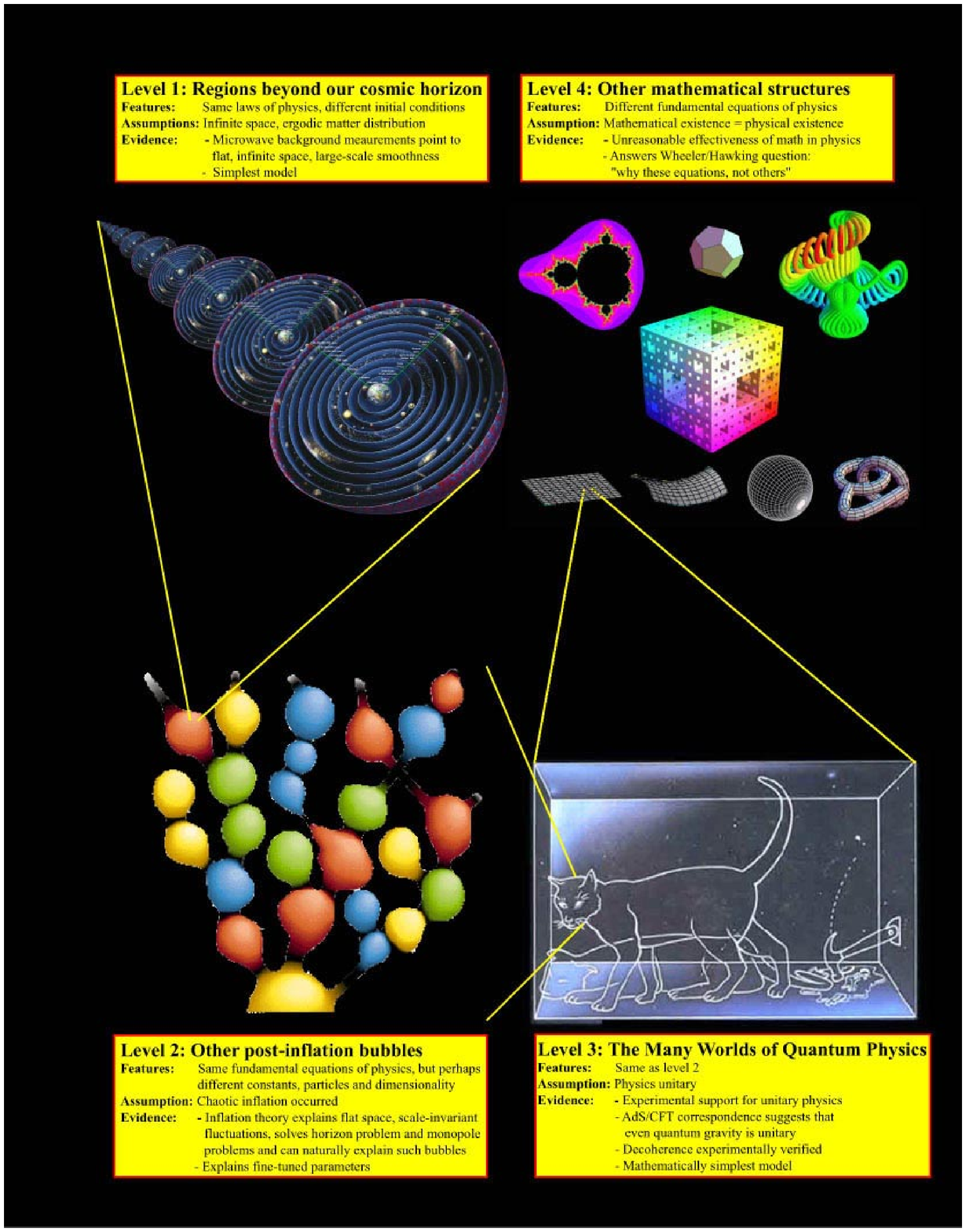}}}}
\end{figure}
\setcounter{figure}{1}
\clearpage

\section{Level I: Regions beyond our cosmic horizon}

Let us return to your distant twin.
If space is infinite and the distribution of matter is sufficiently uniform on 
large scales, then even the most unlikely events must take place somewhere.
In particular, there are infinitely many other inhabited planets, including 
not just one but infinitely many with
people with the same appearance, name and memories as you.
Indeed, there are infinitely many other regions the size of our observable universe,
where every possible cosmic history is played out. This is the Level I multiverse.

\subsection{Evidence for Level I parallel universes}

Although the implications may seem crazy and counter-intuitive, this 
spatially infinite cosmological model is in fact the simplest and most popular one
on the market today. It is part of the cosmological concordance model, 
which agrees with all current observational evidence and is 
used as the basis for most calculations and simulations presented at cosmology conferences.
In contrast, alternatives such as a fractal universe, a closed universe and a multiply connected 
universe have been seriously challenged by observations.
Yet the Level I multiverse idea has been controversial (indeed, an assertion along these lines
was one of the heresies for which the Vatican had Giordano Bruno burned 
at the stake in 1600\footnote{Bruno's ideas have since been elaborated by, \eg,
\cite{Brundrit79,Garriga01}, all of whom have thus far avoided the stake.}), 
so let us review the status of the two assumptions
(infinite space and ``sufficiently uniform'' distribution). 
 
How large is space? Observationally, the lower bound has grown dramatically 
(\fig{SizeFig}) with no indication of an upper bound.
We all accept the existence of things that we cannot see but could see
if we moved or waited, like ships beyond the horizon. 
Objects beyond cosmic horizon have similar status, since the observable universe 
grows by a light-year every year as light from further away has time to 
reach us\footnote{If the cosmic expansion continues to accelerate (currently an open question),
the observable universe will eventually stop growing.
}. 
If anything, the Level I multiverse sounds trivially obvious. How could space not be
infinite? Is there a sign somewhere saying "Space Ends Here--Mind the Gap"? If so,
what lies beyond it? In fact, Einstein's theory of gravity calls this intuition into
question. Space could be finite if it has a convex curvature or an unusual topology
(that is, interconnectedness). A spherical, doughnut-shaped or pretzel-shaped
universe would have a limited volume and no edges. The cosmic microwave background
radiation allows sensitive tests of such scenarios. 
So far, however, the evidence is against them. Infinite models fit the
data, and strong limits have been placed on the alternatives 
\cite{smalluniverse,Shapiro06}.
In addition, 
a spatially infinite universe is a generic prediction of the cosmological theory of 
inflation \cite{Garriga01}, so
the striking successes of inflation listed below therefore lend further support
to the idea that space is after all infinite just as we learned in school.

Another loophole is that space is infinite but matter is confined to a finite region around us -- the
historically popular ``island universe" model. In a variant on this model, matter thins out on large scales in a
fractal pattern. In both cases, almost all universes in the Level I multiverse would be empty and dead. But
recent observations of the three-dimensional galaxy distribution and the microwave background have shown that
the arrangement of matter gives way to dull uniformity on large scales, with no coherent structures larger than
about $10^{24}$ meters. Assuming that this pattern continues, space beyond our observable universe teems with
galaxies, stars and planets.

\begin{figure}[pbt]
\vskip-2.5cm
\caption{Although an infinite universe has always been a possibility, the lower
limit on the size of our universe has kept growing.
}
\label{SizeFig}
\end{figure}

\subsection{What are Level I parallel universes like?}
\label{ErgodicitySec}

The physics description of the world is traditionally split into two parts:
initial conditions and laws of physics specifying how the initial conditions evolve.
Observers living in parallel universes at Level I observe the exact same laws of physics as we do,
but with different initial conditions than those in our Hubble volume.
The currently favored theory is that the initial conditions
(the densities and motions of different types of matter early on)
were created by quantum fluctuations during the inflation epoch (see section 3).
This quantum mechanism generates initial conditions that are for all practical 
purposes random, producing density fluctuations described by what mathematicians
call an ergodic random field.
{\it Ergodic} means that if you imagine 
generating an ensemble of universes, each with its own random initial conditions,
then the probability distribution of outcomes in a given volume
is identical to the distribution that you get by sampling different volumes in a single universe.
In other words, it means that everything that could in principle have happened here 
did in fact happen somewhere else.

Inflation in fact generates all possible initial conditions with non-zero probability, 
the most likely ones
being almost uniform with fluctuations at the $10^{-5}$ level that are amplified by
gravitational clustering to form galaxies, stars, planets and other structures.
This means both that pretty much all imaginable matter configurations occur in some Hubble 
volume far away, and also that we should expect our own Hubble volume to be a
fairly typical one --- at least typical among those that contain observers.
A crude estimate suggests that the closest identical copy of you is about $\sim10^{10^{29}}$m away. 
About $\sim 10^{10^{91}}$m away, there should be a sphere of radius 
100 light-years identical to the one centered here, so all perceptions that we have
during the next century will be identical to those of our counterparts over there.
About $\sim 10^{10^{115}}$m away, there should be an entire Hubble volume identical to 
ours.\footnote{This 
is an extremely conservative estimate, simply counting all possible quantum 
states that a Hubble volume can have that are no hotter than $10^8$K.  
$10^{115}$ is roughly the number of protons that the Pauli exclusion principle
would allow you to pack into a Hubble volume at this temperature
(our own Hubble volume contains only about $10^{80}$ protons).
Each of these $10^{115}$ slots can be either occupied or unoccupied, giving 
$N=2^{10^{115}}\sim 10^{10^{115}}$ possibilities, 
so the expected distance to the nearest identical Hubble volume is 
$N^{1/3}\sim 10^{10^{115}}$ Hubble radii $\sim 10^{10^{115}}$ meters.
Your nearest copy is likely to 
be much closer than $10^{10^{29}}$ meters, since the planet formation and evolutionary processes 
that have tipped the odds in your favor are at work everywhere. There are probably at 
least $10^{20}$ habitable planets in our own Hubble volume alone. 
}

\subsection{How derive probabilities from a causal theory?}
\label{ProbabilityOriginSec}

Let us now turn to worry~4 and worry~5 in Table~1.
The Level I multiverse raises an interesting philosophical point: 
you would not be able to compute your own future even if you had complete knowledge
of the entire state of the cosmos! 
The reason is that there is no way for you to determine 
which of these copies is ``you'' (they all feel that they are). Yet their lives will typically 
begin to differ eventually, so the best you can do is predict probabilities for what you will
observe, corresponding to the fractions of these observers that experience different things.
This kills the traditional notion of determinism even without invoking quantum mechanics. 

However, perhaps it is a uniquely quantum-mechanical phenomenon that you can end up with subjective indeterminism even if only a single you exists to start with? 
No, because we can create the same phenomenon in the following simple gedanken experiment involving only classical physics, without even requiring any sort of multiverse (not even Level I).
You are told that you will be sedated, that a perfect clone of you will be constructed (including your memories), and that the two yous will be woken up by a bell at the same time the next morning in two identical-looking rooms. The rooms are numbered 0 and 1, and these numbers are printed on a sign outside the door. When asked by the anesthesiologist to place a bet on where you will wake up, you realize that you have to give her 50-50 odds, because someone feeling that they are you will wake up in both. When you awaken, you realize that you'd still give 50-50 odds, because even if you knew the position of every atom in the universe, you still couldn't know which of the two yous is the one having your current subjective experience.
When you go outside, the room number you read will therefore feel just like a random number to you.

Now suppose instead that you were told that this experiment would be repeated 10 more times, resulting in a total of $2^{10}$ clones in $1024$ identical rooms which have their numbers written out in binary. When asked to place bets on your room number, you assign an equal probability for all of them. However, you can give more interesting odds on what fraction of the ten binary digits on your door will be zeros, knowing from the binomial theorem that it's $50\%$ for $\left(10\atop 5\right)=254$ yous, 20\% for $\left(10\atop 1\right)=45$ yous, etc. You can therefore say that you will probably see a random-looking string of zeroes and ones on your door, with an 89\% chance that the fraction of ones will be between 30\% and 70\%.
This conclusion is exactly the same as you would draw if you instead assumed that there was only one of you, and that you would be placed in a random room. Or that there was only one one you and one room, whose 10 digits were each generated randomly with 50\% probability for both 0 and 1.

In Everett's MWI, probability appears from randomness in exactly the same way if the branches have equal amplitude: one you evolves into more than one through deterministic Hamiltonian dynamics as in Equation~(2) with $\alpha=\beta=2^{-1/2}$.
The only difference is what the physical nature of the cloning process is. In our example above, another difference is that the hospital guests can meet and verify the existence of their clones, whereas quantum clones cannot because of decoherence --- however, the hospital experiment could easily be modified to have this property too, say by keeping the rooms locked forever or shipping the clones off into deep space without radios.
It is therefore observer cloning that is the crux, not what physics is involved in the cloning process. You need to end up with more than one post-experiment you with different recent experiences, but having identical memories from before the experiment.


In summary, although these classical parallels have not ended the debates over probability in the Everett picture,
as evidenced by the continuing controversy over whether probability makes sense in the many worlds interpretation
(this volume, Part 3, Part 4), they do show that worries~4 and~5 appear already in classical physics.
That is, whenever there are multiple observers with identical memories of
what happened before a certain point but differing afterwards, these observers will perceive
apparent randomness even if the evolution of their universe is completely deterministic. Whenever
an observer is cloned, she will perceive something completely indistinguishable from true
randomness. Since both of these phenomena can be realized without quantum mechanics, apparent
causality breakdown and randomness are therefore not quantum-specific. Unitary quantum mechanics
has these attributes simply because it routinely creates observer
cloning when an instability rapidly amplifies microsuperpositions into macrosuperpositions, 
while decoherence ensures --- effectively --- that the doors between the clones are kept locked forever. 
Examples of such instabilities include most quantum measurements, Schr{\"o}dinger's cat experiment and, probably, 
certain snap decision processes in the brain.

\subsection{How a multiverse theory can be tested and falsified}
\label{TestingSec}

Is a multiverse theory one of metaphysics rather than physics?
This is the concern listed as worry~1 in Table~1. 
As emphasized by Karl Popper, the distinction between the two is whether the theory
is empirically testable and falsifiable.
Containing unobservable entities does clearly {\it not} per se make a theory non-testable.
For instance, a theory stating that there are 666 parallel universes, all of which are devoid of oxygen makes the testable prediction that we should observe no oxygen here, and is therefore ruled out by observation.

As a more serious example, the Level I multiverse framework is routinely used 
to rule out theories in modern cosmology, although this is rarely spelled out explicitly. 
For instance, cosmic microwave background (CMB) observations have recently 
shown that space has almost no curvature.
Hot and cold spots in CMB maps have a characteristic 
size that depends on the curvature of space, and the observed spots appear
too large to be consistent with the previously popular ``open universe'' model.
However, the average spot size randomly varies slightly from one Hubble 
volume to another, so it is important to be statistically rigorous.
When cosmologists say that the open universe model is ruled out at 
99.9\% confidence,
they really mean that if the open universe model were true,
then fewer than one out of every thousand Hubble volumes would show CMB spots as
large as those we observe --- therefore the entire model with frits entire Level I multiverse of infinitely many 
Hubble volumes is ruled out, even though we have of course only 
mapped the CMB in our own particular Hubble volume.

A related issue is worry~8 in Table~1: how does one judge evidence for/against a multiverse theory, if some small fraction of the observers get fooled by unusual data?
For example, of a Stern Gerlach apparatus is used to measure the spin in the $z$-direction of 10000 particles prepared with their spin in the $x$-direction, most of the $2^{10000}$ resulting observers will observe a random looking sequence with about 50\% spin-up, but one of them will be unlucky enough to measure spin up every time and mistakenly conclude that quantum mechanics is incorrect.

This issue clearly has nothing to do with quantum mechanics {\it per se}, since it also occurs in our last hospital example from \Sec{ProbabilityOriginSec}. Suppose the 1024 clones are all considering the hypothesis that what happened to them is indeed the cloning 
experiment as described in \Sec{ProbabilityOriginSec}, trying to decide whether to believe it or not. They all observe their room numbers, and most of them find it looking like a random sequence of zeros and ones, consistent with the hypothesis. However, one of the clones observes the room number "0000000000", and declares that the hypothesis has been ruled out at 99.9\% confidence, because if the hypothesis were true, the probability of finding oneself in the very first room is only $1/1024$.
Similar issues also tormented some of the pioneers of classical statistic mechanics: in the grand ensemble at the heart of the theory, there would always be some totally confused observers who repeatedly saw eggs unbreak and cups of water spontaneously separate into steam and ice.

When they occur in examples not involving quantum mechanics, these issues are generally considered resolved, merely exemplifying what confidence levels are all about. If anybody in any context says that she has ruled something out at 99.9\% confidence, she means that there is a 1 in 1000 chance that she has been fooled. Whenever there is any form of randomness, either ontologically fundamental as in the Copenhagen interpretation, apparent as in the MWI or merely epistemological (reflecting our inability to predict detector noise, say), there is a risk that we get fooled by fluke data. In most cases, we can reduce this risk as much as we want by performing more measurements, but in some cases we cannot, say when measuring the large-scale power spectrum in the cosmic microwave background, where further measurements would only help if we could perform them in outside of our cosmic horizon volume, \ie, in Level I parallel universes.

The take-home message from this section is that the MWI and indeed any multiverse theories {\it can} be 
tested and falsified, but only if they predict what the ensemble of parallel universes is and specify  a probability distribution (or more generally what mathematicians call a {\it measure})
over it. 
This measure problem can be quite serious and is still unsolved for some multiverse theories 
(see \cite{inflation,Garriga05,Easther05,Aguirre06,Bousso06,Page08} for recent reviews),
but is solved for both statistical mechanics and for quantum mechanics in a finite space.

\section{Level II: Other post-inflation bubbles}

If you felt that the Level I multiverse was large and hard to stomach,
try imagining an infinite set of distinct ones
(each symbolized by a bubble in 
Figure 1), some perhaps with different
dimensionality and different physical constants.
This is what is predicted by most 
currently popular models of inflation, and we will refer to
it as the Level II multiverse. 
These other domains are more than infinitely far away in the sense
that you would never get there even if you traveled at the speed of light forever.
The reason is that the space between our Level I multiverse and its neighbors
is still undergoing inflation, which keeps stretching it out 
and creating more volume faster than you can travel through it.
In contrast, you could travel to an arbitrarily distant Level I universe if you were patient 
and the cosmic expansion decelerates.\footnote{
Astronomical evidence suggests that the cosmic expansion is currently accelerating.
If this acceleration continues, then even the level I parallel universes will remain forever separate, 
with the intervening space stretching faster than light can travel through it.
The jury is still out, however, with popular models predicting that the universe will eventually 
stop accelerating and perhaps even recollapse.
}

\subsection{Evidence for Level II parallel universes}

Inflation is an extension of the big bang theory and ties up many of its loose ends, such as why
the universe is so big, so uniform and so flat. An almost exponentially rapid stretching of space long ago can explain all these and other attributes in one fell swoop (see reviews \cite{Linde94,GuthKaiser05}.
Such stretching is predicted by a wide class of theories of elementary particles, and all available
evidence bears it out. Much of space is stretching and will continue doing so forever, 
but some regions of space stop inflating and form
distinct bubbles, like gas pockets in a loaf of rising bread. Infinitely many such bubbles emerge
(Figure 1, lower left, with time increasing upwards).
Each is an embryonic Level I multiverse: infinite in size\footnote{Surprisingly, it has been shown that inflation can produce 
an infinite Level I multiverse even in a bubble of finite spatial volume, thanks
to an effect whereby the spatial directions of spacetime curve towards the (infinite) 
time direction \cite{BucherSpergel999}. 
} and filled with matter deposited by the energy field that drove inflation. 
Recent cosmological measurements have confirmed two key predictions of inflation: that space 
has negligible curvature and that the clumpiness in the cosmic matter distribution used to be approximately scale invariant.

The prevailing view is that the physics we observe today is
merely a low-energy limit of a more general theory that 
manifests itself at extremely high temperatures. 
For example, this underlying fundamental theory
may be 10-dimensional, supersymmetric and involving a grand unification 
of the four fundamental forces of nature.
A common feature in such theories is that the potential energy of the
field(s) relevant to inflation has many different minima 
(sometimes called ``metastable vacuum states''), and ending up in different 
minima corresponds to different effective laws of physics for our low-energy world.
For instance, all but three spatial dimensions could be curled up (``compactified'') on a tiny scale, 
resulting in an effectively three-dimensional space like ours, or 
fewer could curl up leaving a 5-dimensional space.
Quantum fluctuations during inflation can therefore cause different post-inflation bubbles in
the Level II multiverse 
to end up with different effective laws of physics in different bubbles
--- say different dimensionality or different types of elementary particles, 
like two rather than three generations of quarks.

In addition to such discrete properties as dimensionality and particle content,
our universe is characterized by a set of dimensionless 
numbers known as {\it physical constants}. Examples include the electron/proton 
mass ratio $m_p/m_e\approx 1836$ and the cosmological constant, which appears to be about
$10^{-123}$ in so-called Planck units. There are models where also such 
non-integer parameters can vary from one post-inflationary 
bubble to another.\footnote{
Although the fundamental equations of physics are the same throughout the Level II multiverse, the 
approximate effective equations governing the low-energy world that we observe will differ.
For instance, moving from a three-dimensional to a four-dimensional (non-compactified) space 
changes the observed gravitational force equation from an inverse square law to an inverse cube law.
Likewise, breaking the underlying symmetries of particle physics differently will change the lineup 
of elementary particles and the effective equations that describe them.
However, we will reserve the terms ``different equations'' and ``different laws of physics''
for the Level IV multiverse, where it is the fundamental rather than effective equations that change.
}
In summary, the Level II multiverse is likely to be more diverse than the Level I multiverse,
containing domains where not only the initial conditions differ,
but perhaps the dimensionality, the elementary particles and
the physical constants differ as well.

This is currently a very active research area.
The possibility of a string theory ``landscape'' \cite{BoussoPolchinski00,Susskind03},
where the above-mentioned potential has perhaps $10^{500}$ different minima, may offer a specific realization of
the Level II multiverse which would in turn have four sub-levels of increasing diversity:
{\bf IId:} different ways in which space can be compactified, which can allow both different effective
dimensionality and different symmetries/elementary articles (corresponding to different 
topology of the curled up extra dimensions).
{\bf IIc:} different ``fluxes'' (generalized magnetic fields) that stabilize the extra dimensions (this sublevel is where the largest 
number of choices enter, perhaps $10^{500}$).
{\bf IIb:} once these two choices have been made, there may be a handful of different minima in the effective supergravity potential.
{\bf IIa:} the same minimum and effective laws of physics can be realized in a many different post-inflationary bubbles, each 
constituting a Level I multiverse.

Before moving on, let us briefly comment on a few closely related multiverse notions.
First of all, if one Level II multiverse can exist, eternally self-reproducing in a fractal pattern,
then there may well be infinitely many other Level II multiverses that are completely disconnected.
However, this variant appears to be untestable, since it would neither add any qualitatively different worlds 
nor alter the probability distribution for their properties.
All possible initial conditions and symmetry breakings
are already realized within each one.

An idea proposed by Tolman and Wheeler and recently elaborated by 
Steinhardt \& Turok \cite{SteinhardtTurok02} is that the (Level I) multiverse is
cyclic, going through an infinite series of Big Bangs.
If it exists, the ensemble of such incarnations would also form
a multiverse, arguably with a diversity similar to that of Level II.

An idea proposed by Smolin \cite{Smolin97} involves an ensemble similar in diversity to that of Level II, but 
mutating and sprouting new universes through black holes rather than during inflation.
This predicts a form of a natural selection favoring universes with maximal black hole production.

In braneworld scenarios, another 3-dimensional world could be quite
literally parallel to ours, merely offset in a higher dimension. However, it is unclear
whether such a world (``brane'') deserves be be called a parallel universe separate from our own, since 
we may be able to interact with it gravitationally much as we do with dark matter.


\subsection{Fine-tuning and selection effects}
\label{TuningSec}

Although we cannot interact with other Level II parallel universes, cosmologists can infer their
presence indirectly, because their existence can account for unexplained coincidences in our universe.
To give an analogy, suppose you check into a hotel, are assigned room 1967 and note that this is the
year you were born. What a coincidence, you say. After a moment of reflection, however, you conclude
that this is not so surprising after all. The hotel has hundreds of rooms, and you would not have been
having these thoughts in the first place if you had been assigned one with a number that meant nothing
to you. The lesson is that even if you knew nothing about hotels, you could infer the existence of other
hotel rooms to explain the coincidence. 

As a more pertinent example, consider the mass of the sun. The mass of a star determines its luminosity,
and using basic physics, one can compute that life as we know it on Earth is possible only if the sun's
mass falls into the narrow range between $1.6\times 10^{30}\kg$ and $2.4\times 10^{30}\kg$.
Otherwise Earth's climate
would be colder than that of present-day Mars or hotter than that of present-day Venus. The measured
solar mass is $M\sim 2.0\times 10^{30}\kg$. At first glance, this apparent coincidence of the habitable and
observed mass values appears to be a wild stroke of luck. Stellar masses run from $10^{29}$ 
to $10^{32}\kg$, so if the sun acquired its mass at random, it had only a small chance of falling into the
habitable range. But just as in the hotel example, one can explain this apparent coincidence by
postulating an ensemble (in this case, a number of planetary systems) and a selection effect (the fact
that we must find ourselves living on a habitable planet). Such observer-related selection effects are
referred to as ``anthropic'' \cite{Carter74}, and although the ``A-word'' is notorious for triggering controversy,
physicists broadly agree that these selection effects cannot be neglected when testing fundamental
theories. In this weak sense, the anthropic principle is not optional.

What applies to hotel rooms and planetary systems applies to parallel universes. Most, if not all, of the attributes set by
symmetry breaking appear to be fine-tuned. Changing their values by modest amounts would have resulted in a qualitatively
different universe--one in which we probably would not exist. If protons were $0.2\%$ heavier, they could decay into
neutrons, destabilizing atoms. If the electromagnetic force were $4\%$ weaker, there would be no hydrogen and no normal
stars. If the weak interaction were much weaker, hydrogen would not exist; if it were much stronger, supernovae would fail
to seed interstellar space with heavy elements. If the cosmological constant were much larger, the universe would have
blown itself apart before galaxies could form.
Indeed, most if not all the parameters affecting low-energy physics appear fine-tuned
at some level, in the sense that changing them by modest amounts results in 
a qualitatively different universe.

Although the degree of fine-tuning is still debated
(see \cite{BarrowTipler,toe,Hogan00,axions}) for more technical reviews),
these examples suggest the existence of parallel universes with other
values of some physical constants. 
The existence of a Level II multiverse implies that physicists will never be able to determine the values of all
physical constants from first principles. Rather, they will merely compute probability distributions for what they 
should expect to find, taking selection effects into account. 
The result should be as generic as is consistent with our existence.


\begin{figure}[pbt]
\centerline{{\vbox{\epsfxsize=8.7cm\epsfbox{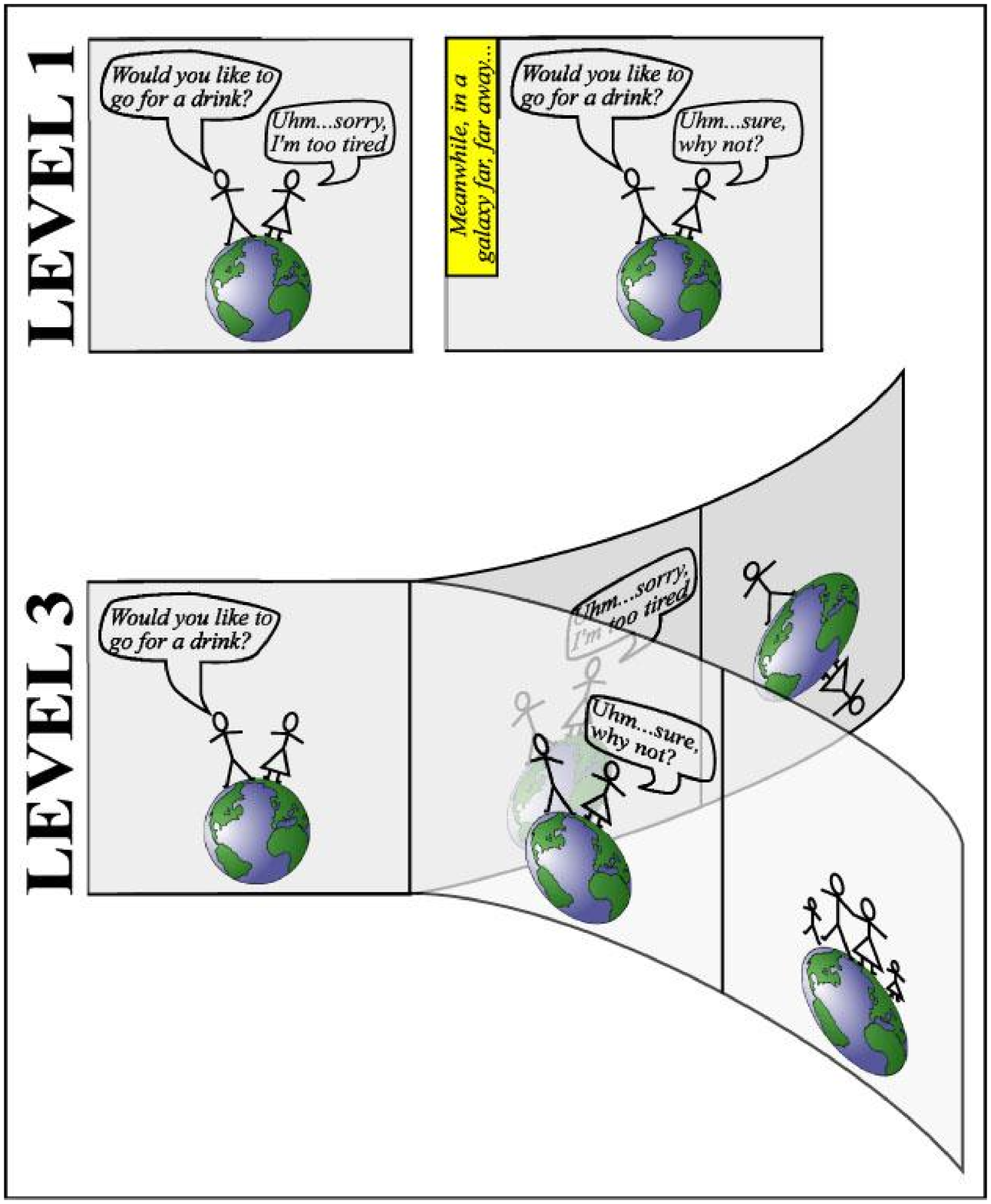}}}}
\smallskip
\caption{
Difference between Level I and Level III.
Whereas Level I parallel universes are far away in space, those of Level III
are even right here, with quantum events causing classical reality to split and diverge into 
parallel storylines. Yet Level III adds no new storylines beyond levels 1 or 2.
}
\label{CartoonFig}
\end{figure}

\section{Level III: The many worlds of quantum physics}

If Everett was correct and physics is unitary, then there is a third type of parallel worlds that are 
not far away but in a sense right here.
The universe keeps branching into parallel universes 
as in the cartoon (\fig{CartoonFig}, bottom): whenever a quantum event appears to have a random outcome, 
all outcomes in fact occur, one in each branch.
This is the Level III multiverse. Although more debated and controversial than 
Level I and Level II, we will see that, surprisingly, 
this level adds no new types of universes.

Since the volume to which this chapter belongs discusses the MWI in great detail, we will summarize the key points only very briefly.
Everett's MWI is simply standard quantum mechanics with the collapse postulate removed, so that the Schr\"odinger equation holds without exception (\Sec{MWIdefSec}). From this, the following conclusions can be derived:
\begin{enumerate}
\item Microsuperpositions (say of an atom going through two slits at the same time) are inevitable (the Heisenberg Uncertainty principle).
\item Macrosuperpositions (say of a cat being dead and alive) are also perfectly legitimate quantum states.
\item Processes occur that amplify microsuperpositions into macrosuperpositions (spontaneous symmetry breaking, Schr{\"o}dinger's cat, and quantum measurements being three examples).
\item The superposition of a single macroscopic object tends to spread to all other interacting objects, eventually engulfing our entire universe. 
\item Decoherence makes most macrosuperpositions for all practical purposes unobservable.
\item Decoherence calculations can determine which quantities appear approximately classical.
\end{enumerate}
There is consensus in the physics community that both double-slit interference and the process of decoherence have been experimentally observed, showing the predicted behavior. 
Conclusions 1, 2, 3 and 4 together imply that astronomically large macrosuperpositions occur. These are Everett's parallel 
universes.\footnote{Note that to avoid creating macrosuperpositions, it is insufficient to abandon unitarity. Rather, it is {\it symmetry} that must be abandoned.
For example, any theory where the wavefunction of a system evolves deterministically 
(even if according to another rule than the Schr{\"o}dinger equation) will evolve a perfectly sharp needle balanced on its tip into a superposition of needles pointing in 
macroscopically different directions unless the threory explicitly violates rotational symmetry. 
If the theory does violate this symmetry and ``collapses'' the wavefunction, then there are two interesting possibilities: 
either the symmetry is broken early on while the superposition is still microscopic and unobservable (in which case the collapse process has nothing to do with measurement),
or the symmetry is broken later on when the superposition is macroscopic (in which case local energy conservation is seriously violated by 
abruptly moving the center-of-mass by a macroscopic amount --- even if the mass transfer is not superluminal, it would have to be fast enough to 
involve kinetic energy greatly exceeding the natural energy scale of the problem). 
If this experiment or Schr{\"o}dinger's cat experiment were performed in a sealed free-falling box, the environment outside the box would learn how the needle had fallen or whether the
cat had died from the altered gravitational field outside the box (and perhaps also from recoil motion of the box), causing decoherence. However, this complication can in principle 
be eliminated by keeping the moving parts spherically symmetric at all times. For example, if a metal sphere full of hydrogen contains a smaller sphere at its center full of oxygen at 
the same pressure which is opened if an atom decays (after which diffusion would mix the gases), the resulting superposition of two macroscopically different density distributions 
would leave all external fields unaffected.
}
Worry~10 in Table~1 is addressed by 5, and worry~11 is addressed by 6 as reviewed in \cite{ZehBook,quantum,Zurek09}. 
It should be borne in mind that these two worries remained serious open problems when Everett first published his work, since decoherence was only discovered in 1970 \cite{Zeh70}.

\subsection{What are Level III parallel universes like?}

Everett's many-worlds interpretation has been boggling minds inside and outside physics
for more than four decades. But the theory becomes easier to grasp when one distinguishes
between two ways of viewing a physical theory: the outside view of a physicist studying
its mathematical equations, like a bird surveying a landscape from high above it, and the
inside view of an observer living in the world described by the equations, like a frog
living in the landscape surveyed by the bird.

From the bird perspective, the Level III multiverse is simple. There is only one wave
function. It evolves smoothly and deterministically over time without any kind of
splitting or parallelism. The abstract quantum world described by this evolving wave
function contains within it a vast number of parallel classical story lines, continuously
splitting and merging, as well as a number of quantum phenomena that lack a classical
description. From their frog perspective, observers perceive only a tiny fraction of this
full reality. They can view their own Level I universe, but the process of
decoherence \cite{Zeh70,ZehBook} --- which mimics wave function collapse while preserving unitarity--prevents
them from seeing Level III parallel copies of themselves. 

Whenever observers are asked a question, make a snap decision and give an answer, quantum
effects in their brains lead to a superposition of outcomes, such as ``Continue reading
the article'' and ``Put down the article''. From the bird perspective, the act of making a
decision causes a person to split into multiple copies: one who keeps on reading and one
who doesn't. From their frog perspective, however, each of these alter egos is unaware of
the others and notices the branching merely as a slight randomness: a certain probability
of continuing to read or not. 

As strange as this may sound, the exact same situation occurs even in the Level I
multiverse. You have evidently decided to keep on reading the article, but one of your
alter egos in a distant galaxy put down the magazine after the first paragraph. The only
difference between Level I and Level III is where your doppelga{\"a}ngers reside. In Level I
they live elsewhere in good old three-dimensional space. In Level III they live on
another quantum branch in infinite-dimensional Hilbert space (\fig{CartoonFig}). 

\subsection{Level III parallel universes: evidence \& implications}

The existence of Level III depends on one crucial assumption: that the time evolution of the wave function is unitary. So
far experimenters have encountered no departures from unitarity. In the past few decades they have confirmed unitarity for
ever larger systems, including carbon 60 buckyball molecules and kilometer-long optical fibers. On the theoretical side,
the case for unitarity has been bolstered by the discovery of decoherence (see \cite{quantum} for a popular review). 
Some theorists who work on quantum gravity have
questioned unitarity; one concern is that evaporating black holes might destroy information, which would be a nonunitary
process. But a recent breakthrough in string theory known as AdS/CFT correspondence suggests that even quantum gravity is
unitary. If so, black holes do not destroy information but merely transmit it elsewhere.

If physics is unitary, then the standard picture of how quantum fluctuations operated early in the big bang must change.
These fluctuations did not generate initial conditions at random. Rather they generated a quantum superposition of all
possible initial conditions, which coexisted simultaneously. Decoherence then caused these initial conditions to behave
classically in separate quantum branches. Here is the crucial point: the distribution of outcomes on different quantum
branches in a given Hubble volume (Level III) is identical to the distribution of outcomes in different Hubble volumes
within a single quantum branch (Level I). This property of the quantum fluctuations is known in statistical mechanics as
ergodicity. 

The same reasoning applies to Level II. The process of symmetry breaking did not produce a unique outcome but rather a
superposition of all outcomes, which rapidly went their separate ways. So if physical constants, spacetime dimensionality
and so on can vary among parallel quantum branches at Level III, then they will also vary among parallel universes at
Level II. 

In other words, the Level III multiverse adds nothing new beyond Level I and Level II, just more indistinguishable copies
of the same universes--the same old story lines playing out again and again in other quantum branches. The passionate
debate about Everett's theory therefore seems to be ending in a grand anticlimax, with the discovery of less controversial
multiverses (Levels I and II) that are equally large.

\subsection{The unequal probability worry}

Let us now turn to worry~6 in Table 1: how to compute the apparent probabilities from the wave function amplitudes when they are not all equal and the wave-function collapse postulate has been dropped from the theory.
Since a single approximately classical state often evolves into a superposition of macroscopically different states that rapidly decohere as discussed above, it is obvious that observers will experience apparent randomness just as in our hospital examples from \Sec{ProbabilityOriginSec}. However, why is it that these probabilities correspond to the square modulus of the wave function amplitudes (the so-called Born rule)? For example, in \eq{SplitEq}, why is the apparent probability for a happy observer equal to $|\alpha^2|$ rather than some other real-valued function of $\alpha$, say $|\alpha|^4$?

There are a number of 
arguments that suggest that it must be this way. For example, one could argue that the sum of the probabilities should be conserved (so that it can be normalized to 1 once and for all), and $\int|\phi|^2$ is the only functional of $\psi$ that is conserved under arbitrary unitary evolution, by the very definition of unitarity. In other words, the business about the squaring comes straight from the Hilbert-space structure of quantum mechanics, whereby the inner product defines an $L_2$ norm but no other norms.


Other arguments to this end have been proposed, based on information theory \cite{EverettBook},
decision theory \cite{Deutsch99} and other approaches \cite{Caves02,Saunders02,Zurek09}.
But many authors have expressed a deeper concern about whether probability in the usual sense even makes sense in MWI
(often focused around some combination of worries~4 and~5). 
To this end, arguments have been proposed based on
Savage's approach: whatever intelligent observers actually believe, they will behave as though
ascribing subjective probabilities to outcomes --- probabilities which, as \cite{Deutsch99,Wallace02,Wallace03} showed, match the Born rule.
A rigorous mathematical treatment of this is given by Wallace in Chapter of this volume.

At the extensive debates about this issue at the  ``Everett @ 50'' conference at the Perimiter Institute in 2007, it was clear that these purported Born Rule derivations were still controversial. Interestingly, the entire controversy centered around the equal-probability case
(say $\alpha=\beta$ in \eq{SplitEq}), \ie, getting probabilities in the first place (worries~4 and~5 in Table~1). In contrast, the notion that this can be generalized to arbitrary amplitudes (worry~6 in Table~1) was fairly uncontroversial.
In summary, worry~6 is the first one in Table~1 which is truly specific to quantum mechanics, but addressing it if 
worries~4 and~5 have been settled is arguably a solved problem.

\subsection{Does the state describe the world or my knowledge of it?}

A quantum state can be mathematically described by a density matrix.
But what does this density matrix really describe? The state of the universe or your state of knowledge about it?
This issue, listed as worry~7 in Table~1, is as old as quantum mechanics itself and still divides the physics community.

The Everett postulate implies a clear answer to it: both!
On one hand, the entire universe has a quantum state which corresponds to a wavefunction, or to a density matrix if the state is mixed. Let us call this the {\it ontological} quantum state. On the other hand, our state of knowledge of the universe is described by a lower-dimensional density matrix for those degrees of freedom that we are interested in, both conditioned on what we already know (limiting to those branches that we could be on --- what Everett termed the ``relative state'') and partial-traced over those degrees of freedom that we know nothing about. I will refer to this as the {\it epistemological} quantum state, bearing in mind that it differs from one observer to the another --- both from a colleague in this branch of the wavefunction and from yourself in another branch\footnote{Whereas the ontological state might be pure and hence describably by a wave function, the epistemological state is generically mixed and cannot be described by a wavefunction, only by a density matrix. This was pointed out already by Schr{\"o}dinger \cite{Schroedinger35}}.
In other words, the epistemological quantum state is derivable from the ontological quantum state and your subjective observations.
When quantum textbooks refer to ``the'' state, they usually mean the the epistemological state of a system according to you, after you have prepared it in a certain way.
This is further elaborated in \cite{brain}.

The density-matrix aspect of this issue is clearly quantum-specific. However, the dichotomy between objective and subjective descriptions appears in classical statistical mechanics as well: an ensemble of classical worlds can be completely described by a probability distribution in a high-dimensional phase space, whereas the knowledge of the world by an individual observer is described by a probability distribution in a lower-dimensional phase space, again computable by 
conditioning (the classical equivalent of computing a relative state) and 
marginalizing (the classical equivalent of partial tracing).

\subsection{The weirdness worry}
\label{WeirdnessSec}

Despite all the elaborate technical and philosophical worries about the MWI listed in Table~1, many physicists probably find their strongest objection to the MWI not in their brain but in their gut: it simply feels too weird, crazy, 
counter-intuitive and disturbing. 

The complaint about weirdness is aesthetic rather than scientific, and it really makes
sense only in the Aristotelian world view. Yet what did we expect? When we ask a profound
question about the nature of reality, do we not expect an answer that sounds strange?
I personally dismiss this weirdness worry as a failure to 
appreciate Darwinian evolution. Evolution endowed us with intuition only for those aspects of physics that 
had survival value for our distant ancestors, such as the parabolic trajectories of flying rocks. Darwin's 
theory thus makes the testable prediction that whenever we look beyond the human scale, our evolved 
intuition should break down.

We have repeatedly tested this prediction, and the results overwhelmingly support it: our intuition 
breaks down at high speeds where time slows down, on small scales where particles can be in two places at 
once, on large scales where we encounter black holes, and at high temperatures, where colliding particles change identity. To me, an electron colliding with a positron and turning into a Z-boson feels about as intuitive as two colliding cars turning into a cruise ship. The point is that if we dismiss seemingly weird theories out of hand, we risk dismissing the correct theory if we stumble across it.

\subsection{Two world views}

The seemingly endless debate over the interpretation of quantum mechanics is in a sense the
tip of an iceberg.
In the Sci-Fi spoof ``Hitchhiker's Guide to the Galaxy'', the answer 
is discovered to be ``42'', and the hard part is finding the real question.
Questions about parallel universes may seem to be just about as deep as queries about 
reality can get. Yet there is a still deeper underlying question:
there are two tenable but diametrically opposed
paradigms regarding physical reality and the status of mathematics, a dichotomy that  
arguably goes as far back as Plato and Aristotle, and the question is which one is correct.
\begin{itemize}
\item {\bf ARISTOTELIAN PARADIGM:} The subjectively perceived frog perspective is physically real, 
and the bird perspective and all its mathematical language is 
merely a useful approximation.
\item {\bf PLATONIC PARADIGM:} The bird perspective (the mathematical structure) is physically 
real, and the frog perspective and all the human language we use
to describe it is merely a useful approximation for describing
our subjective perceptions.
\end{itemize}
What is more basic --- the frog perspective or the bird perspective?
What is more basic --- human language or mathematical language?
Your answer will determine how you feel about parallel universes.

If you prefer the Aristotelian paradigm, you share worry~3 in Table~1. 
If you prefer the Platonic paradigm, you should find multiverses natural, 
since our feeling that say the Level III multiverse is ``weird'' merely reflects that
the frog and bird perspectives are extremely different. We break the symmetry by calling the 
latter weird because we were all indoctrinated with the Aristotelian 
paradigm as children, long before we even heard of mathematics - the Platonic view is an acquired taste!

In the second (Platonic) case, all of physics is ultimately a mathematics problem, since 
an infinitely intelligent mathematician given the fundamental equations of
the cosmos could in principle
{\it compute} the frog perspective, {\ie}, 
compute what self-aware observers the universe would contain, 
what they would perceive, and what language they would 
invent to describe their perceptions to one another. 
In other words, there is a  ``Theory of Everything" (TOE) 
whose axioms are purely mathematical, since postulates in English 
regarding interpretation would be derivable and thus redundant.
In the Aristotelian paradigm, on the other hand, there can never be a 
TOE, since one is ultimately just explaining
certain verbal statements by other verbal statements ---
this is known as the infinite regress problem \cite{Nozick81}.

In \cite{toe2,toe2newsci}, I have argued that the Platonic paradigm follows logically from the innocuous-sounding {\it External Reality Hypothesis} (ERH) \cite{toe2}: ``there exists an external physical reality completely independent of us humans''. 
More specifically,  \cite{toe2} argues that the ERH implies the {\it Mathematical Universe Hypothesis''} (MUH) that our external physical reality is a mathematical structure. The detailed technical definition of a mathematical structure is not important here; just think of it as a set of abstract entities with relations between them --- familiar examples of mathematical structures include
the integers, a Riemannian manifold, and a Hilbert space.

\section{Level IV: Other mathematical structures}

Suppose you buy the Mathematical Universe Hypothesis and believe that
we simply have not found the correct equations yet, or more rigorously, the correct mathematical structure?
Then an embarrassing question remains, as emphasized by John Archibald
Wheeler: {\it  Why these particular equations, not others?}
\cite{toe2} argues that, when pushed to its extreme, the MUH implies that all mathematical structures correspond to physical universes. Together, these structures 
form the Level IV multiverse, which 
includes all the other levels within it. If there is a particular mathematical structure that is our universe, and 
its properties correspond to our physical laws, then each mathematical structure with different properties is 
its own universe with different laws. The Level IV multiverse is compulsory, since mathematical 
structures are not ``created" and don't exist ``somewhere" --- they just exist. Stephen Hawking once asked, 
``What is it that breathes fire into the equations and makes a universe for them to describe?" In the case of 
the mathematical cosmos, there is no fire-breathing required, since the point is not that a mathematical 
structure describes a universe, but that it is a universe. 

In a famous essay, Wigner \cite{Wigner67} argued that 
``the enormous usefulness of mathematics in the
natural sciences is something bordering on the mysterious", and that
``there is no rational explanation for it".
This argument can be taken as support for the MUH: here
the utility of mathematics for describing the physical
world is a natural consequence of the fact that the latter {\it is}
a mathematical structure, and we are simply uncovering this 
bit by bit.
The various approximations that constitute our current physics theories
are successful because simple mathematical
structures can provide good approximations
of how an observer will perceive more complex mathematical structures.
In other words, our successful theories are
not mathematics approximating physics,
but mathematics approximating mathematics.
Wigner's observation is unlikely to be based on fluke coincidences,
since far more mathematical regularity in nature has been discovered in the 
decades since he made it, including the standard model of particle physics.
Detailed discussions of the Level IV multiverse, what it means and what it predicts are given in \cite{toe,toe2}.

\section{Discussion}
\label{DiscSec}

We have discussed Everett's Many-Worlds Interpretation of quantum mechanics in the context of other 
physics disputes and the three other levels of parallel universes that have been proposed in the literature.
We found that only a small fraction of the usual objections to Everett's theory (summarized in Table~1) are specific to quantum mechanics, and that all of the most controversial issues crop up also in settings that have nothing to do with quantum mechanics.

\subsection{The multiverse hierarchy}

We have seen that scientific theories of parallel universes form a four-level hierarchy,
in which universes become progressively more different from ours. They might have
different initial conditions (Level I), different effective physical laws, 
constants and particles (Level II), or different fundamental physical laws (Level IV). 
It is ironic that Everett's Level III is the one that
has drawn the most fire in the past decades, because it is the only one that adds no
qualitatively new types of universes. 

Whereas the Level I universes join seemlessly, there are clear demarcations
between those within levels II and III caused by 
inflating space and decoherence, respectively. The level IV universes are completely 
disconnected and need to be considered together only 
for predicting your future, since ``you'' may exist in more than one of them.

\subsection{Are parallel universes wasteful?}

A common argument about all forms of parallel universes, including Everett's Level III ones, is that they feel wasteful.
Specifically, the wastefulness worry (\#2 in Table~1) is that multiverse theories
are vulnerable to Occam's razor because they postulate the existence of other worlds that
we can never observe. Why should nature be so wasteful and indulge in such opulence as an
infinity of different worlds? Yet this argument can be turned around to argue for a
multiverse. What precisely would nature be wasting? Certainly not space, mass or
atoms -- the uncontroversial Level I multiverse already contains an infinite amount of all
three, so who cares if nature wastes some more? The real issue here is the apparent
reduction in simplicity. A skeptic worries about all the information necessary to specify
all those unseen worlds. 

But an entire ensemble is often much simpler than one of its members. This principle can
be stated more formally using the notion of algorithmic information content. The
algorithmic information content in a number is, roughly speaking, the length of the
shortest computer program that will produce that number as output. For example, consider
the set of all integers. Which is simpler, the whole set or just one number? Naively, you
might think that a single number is simpler, but the entire set can be generated by quite
a trivial computer program, whereas a single number can be hugely long. Therefore, the
whole set is actually simpler. 

Similarly, the set of all solutions to Einstein's field equations is simpler than a
specific solution. The former is described by a few equations, whereas the latter requires
the specification of vast amounts of initial data on some hypersurface. 
The lesson is that
complexity increases when we restrict our attention to one particular element in an
ensemble, thereby losing the symmetry and simplicity that were inherent in the totality of
all the elements taken together.

In this sense, the higher-level multiverses are simpler. Going from our universe to the
Level I multiverse eliminates the need to specify initial conditions, upgrading to Level
II eliminates the need to specify physical constants, and the Level IV multiverse
eliminates the need to specify anything at all. The opulence of complexity is all in the
subjective perceptions of observers \cite{nihilo} --- the frog perspective. From the bird perspective, the
multiverse could hardly be any simpler. 

A common feature of all four multiverse levels is that the simplest and arguably most
elegant theory involves parallel universes by default. To deny the existence of those
universes, one needs to complicate the theory by adding experimentally unsupported
processes and ad hoc postulates: finite space, wave function collapse, ontological
asymmetry, {\etc} Our judgment therefore comes down to which we find more wasteful and inelegant:
many worlds or many words.

\subsection{Are parallel universes testable}

We have discussed how multiverses are not a theories but predictions of certain theories, and how such theories are falsifiable as long as they also predict something that we can test here in our own universe.
There are ample future prospects for testing and perhaps ruling out these multiverse theories.
In the coming decade, dramatically improved cosmological measurements of the 
microwave background radiation, the large-scale matter distribution, \etc, will test Level I 
by further constraining the curvature and topology of space and will test level II by providing 
stringent tests of inflation.
Progress in both astrophysics and high-energy physics should also clarify the extent to which various
physical constants are fine-tuned, thereby weakening or strengthening the case for Level II.
If the current world-wide effort to build quantum computers succeeds, it will provide 
further evidence for Level III, since such computers are most easily explained as, in essence, 
exploiting the parallelism of the Level III multiverse for parallel 
computation \cite{DeutschBook}.
Conversely, experimental evidence of unitarity violation would rule out Level III.
unifying general relativity and quantum field theory, will shed more light on Level IV. 
Either we will eventually find a mathematical structure matching our universe, or we will 
and have to abandon Level IV.

\subsection{So was Everett right?}

Our conclusions regarding Table~1 do not {\it per se} argue either for or against the MWI, merely clarify what assumptions about physics lead to what conclusions. 
However, all the controversial issues arguably melt away if we accept the {\it External Reality Hypothesis} (ERH) \cite{toe2}: there exists an external physical reality completely independent of us humans. Suppose that this hypothesis is correct. Then the core MWI critique rests on some combination of the following three dubious assumptions.
\begin{enumerate}
\item {\bf Omnivision assumption:} physical reality must be such that at least one observer can in principle observe all of it.
\item {\bf Pedagogical reality assumption:} physical reality must be such that all reasonably informed human observers feel they intuitively understand it.
\item {\bf No-copy assumption:} no physical process can copy observers or create subjectively indistinguishable observers.
\end{enumerate}
1 and 2 appear to be motivated by little more than human h{\"u}bris. 
The omnivision assumption effectively redefines the word ``exists'' to be synonymous with what is observable to us humans.
Of course those who insist on the pedagogical reality assumption will typically have rejected comfortingly familiar childhood notions like Santa Claus, local realism, the Tooth Fairy, and creationism --- but have they really worked hard enough to free themselves from comfortingly familiar notions that are more deeply rooted?
In my personal opinion, our job as scientists is to try to figure out how the world works, not to tell it how to work based on our philosophical preconceptions. 

If the omnivision assumption is false, then there are unobservable things that exist and we live in a multiverse. 
If the pedagogical reality assumption is false, then the weirdness worry (\#12 in Table~1) makes no sense.
If the no-copy assumption is false, then worries~4 and~5 from Table~1 are misguided: observers can perceive apparent randomness even if physical reality is completely deterministic and known. In this case, these fundamental conceptual questions raised by the MWI will arise in physics anyway, independent
of quantum mechanics, and will need to be solved --- indeed, Everett, in providing a coherent and intelligible account of probability even in the face of massive copying, has blazed a trail in showing us how to solve them.

The ERH alone settles worry~9 in Table~1, since what is in the external reality defines what exists.
In summary, if the ERH is correct, then the only outstanding question about the MWI is whether physics is unitary or not. 
So far, experiments have revealed no evidence of unitarity violation, and ongoing and upcoming experiments will test unitarity for dramatically larger systems. 

My guess is that the only issues that worried Hugh Everett were 10 and~11 from Table~1, which are precisely those which were laid to rest by the subsequent discovery of decoherence. Perhaps we will gradually get more used to the weird ways of our cosmos, 
and even find its strangeness to be part of its charm.
In fact, I met Hugh Everett the other day and he told me that he agrees --- but alas not in this particular universe.

\bigskip
{\bf Acknowledgements: }
The author wishes to thank Anthony Aguirre, David Albert, Bryan Eastin, Peter Byrne, 
Olaf Dreyer, Mark Everett, Brian Greene, Alan Guth, Seth Lloyd, George Musser, 
David Raub, Martin Rees, Simon Saunders, Harold Shapiro, Lee Smolin, 
Anton Zeilinger, Wojciech Zurek, Alex Vilenkin and Frank Wilczek 
for stimulating discussions, Simon Saunders for detailed feedback on this manuscript, 
and Adrian Kent, Jonathan Barrett, David Wallace and the Perimeter Institute for hosting a very stimulating meeting on the MWI.

This work was supported by
NSF grants AST-0071213 \& AST-0134999,
NASA grants NAG5-9194 \& NAG5-11099,
a grant from the John Templeton Foundation,
a fellowship from the David and Lucile Packard Foundation
and a Cottrell Scholarship from Research Corporation.


 
 

\end{document}